\documentclass[conference]{IEEEtran}
\usepackage{cite}
\usepackage{amsmath,amssymb,amsfonts}
\usepackage{algorithmic}
\usepackage{graphicx}
\usepackage{textcomp}
\graphicspath{{images/}}
\def\BibTeX{{\rm B\kern-.05em{\sc i\kern-.025em b}\kern-.08em
    T\kern-.1667em\lower.7ex\hbox{E}\kern-.125emX}}
\renewcommand{\baselinestretch}{1}

\begin{document}

\title{Joint Optimization of Application Specific Routing in an Anycast Network\\
\thanks{}
}

\author{Sumit Maheshwari\\
  \textit{WINLAB/ECE, Rutgers University} \\
  \texttt{sumitm@winlab.rutgers.edu}}

\maketitle

\begin{abstract}
Recent developments in the field of Networking have provided opportunities for networks to efficiently cater application specific needs of a user. In this context, a routing path is not only dependent upon the network states but also is calculated in the best interest of an application using the network.  These advanced routing algorithms can exploit application state data to enhance advanced network services such as anycast, edge cloud computing and cyber physical systems (CPS). In this work, we aim to design such a routing algorithm where the router decisions are based upon convex optimization techniques.\\
\end{abstract}

\section{INTRODUCTION}
Next generation of networks are expected to support users in the order of Billion with millisecond latency to all of them. To this end, the optimization carried out at the access level of network have reached to its pinnacle with multitude of optimizations ranging from MIMO, OFDMA etc. [1]. while the network router still take 1 ms to process each L3 packet. In fact, network level optimizations are proposed with techniques such as Software Defined Networking (SDN), Information Centric Networks (ICN) and novel packet processing frameworks such as P4. These techniques rely upon having a centralized controller with global state information available while in real-network, the states are distributed [6],[7] and need time to converge even if dijkstra-like algorithms are used.\linebreak
Literature presents work relating to optimizing network with distributed states as well as joint optimization using convex methods but do not detail how to (a) implement these in live network, (b) utilize distributed states globally and (c) have a system-wide optimization [1-3].

With the advent of anycast mechanism as shown in Fig. 1, a client is able to connect to a \textit{best} server while the definition of \textit{'best'} is fuzzy. In a traditional network, choosing a best server means one with lowest latency while in a modern network the definition may be redefined considering multiple constraints such as energy, application's current status, server load and link cost (i.e. bandwidth and delay). This provides us an opportunity to pursue a joint optimization which not only considers client's latency requirement but also takes acute account of system load and energy. It is to be noted that such an optimization does not hurt the application as not all the  applications require lowest latency and can run (with required Quality of Experience) optimally.

\begin{figure}[thpb]\label{fig:anycast}
\centering
\includegraphics[width=0.5\textwidth]{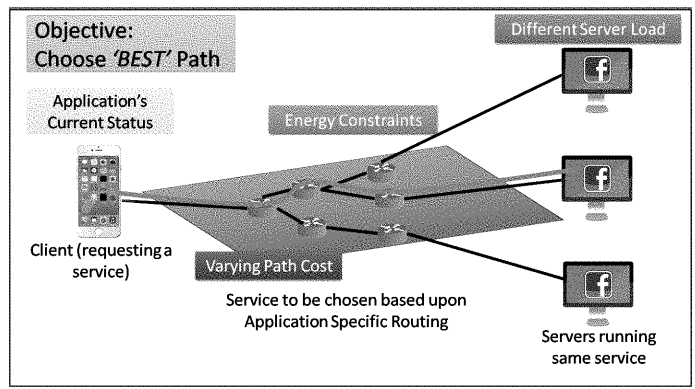}
\caption{Illustration of Application Specific Routing}
\end{figure}

Considering above scenario, we therefore propose the following objective for this work:
\textit{Provide an optimized path rule in a multiuser scenario considering application as well as network state to determine routing}.\\ Please refer Figure \ref{fig:anycast}, where a client is requesting for service A and is being served by best available server not only depending upon the network state but also on user parameters such as (a) Average delay (b) Throughput (c) Energy and (d) Server Load. The network router obtains this information by application pushing these stats after every few milliseconds. The router also has in-network information such as (a) Link Cost (b) Bandwidth (c) Server availability (d) Loss rate and (e) Server Load. Combining these information, a router can best decide the path to serve an application. 

\section{Mathematical Formulation}
In order to understand the problem properly, let us put forward the assumptions before detailing the approach to formulate it mathematically.
\subsection{Assumptions}
In order to provide a simple yet tractable model, we use a centralized controller with following assumptions in use. 
\begin{itemize}
\item{Global information about the network is available at a central controller.}
 \item{Node can get the updated routing information in real-time}
 \item{The energy information can be obtained in real-time from the node or can be calculated using simple models.}
\end{itemize}

\subsection{Notations Used}
The notations used in the work are detailed in this subsection. Consider a network with $G(V,E)\to $ Graph of V nodes and E edges in which $V=\{{{v}_{1}},{{v}_{2}},...,{{v}_{n}})$ is a set of nodes, and
$E=\{{{e}_{1}},{{e}_{2}},...,{{e}_{k}}\}$ is a set of links. At each node,
$b({{e}_{ij}})$ is the bandwidth, $d({{e}_{i}})$ queuing delay and 
$c({{e}_{ij}})$ Energy consumed for sending data from ${{v}_{i}}\to {{v}_{j}}$. Also, for each Path, we have
Path Delay, \[{{d}_{k}}(l)=\sum\limits_{i=1}^{m}{{{d}_{k}}({{e}_{i}})}\] for m nodes in the path.\\
Path Bandwidth: \[{{b}_{k}}(l)=\min {{b}_{k}}({{e}_{ij}})\] \\
Path Energy: \[{{c}_{k}}(l)=\sum\limits_{i=1}^{m}{{{c}_{k}}({{e}_{ij}})}\] \\
Server Load: ${{s}_{k}}(l)$.\\\\
We can understand the rationale behind these as follows: (a) The total path delay is the sum of delay of all links in the path, (b) Path bandwidth is the bottleneck link bandwidth because any node cannot send more data than the bottleneck link, (c) Path energy is the sum of all the energy consumed in the path and (d) Server load is the load observed at the server chosen by the given path \textit{l}. 

\subsection{Problem Formulation}
The problem of finding an joint optimal path from a client to the server thus can be formulated as follows:\\
minimize:     \[\sum\limits_{l=1}^{N}{\alpha d(l)c(l)s(l)b{{(l)}^{-1}}}\] 
subject to:\\
\centerline{$\frac{1}{r}\sum\limits_{i=1}^{r}{{{s}_{i}}(l)}<1$ }
\[\frac{1}{{{C}_{Total}}}\sum\limits_{i=1}^{N}{{{c}_{i}}(l)}\le 1\] 
\centerline{$\frac{{{d}_{i}}(l)}{{{d}_{\max i}}(l)}<1$}
\centerline{${{b}_{\min i}}(l){{b}_{i}}{{(l)}^{-1}}<=1$}
\centerline{${{b}_{\max i}}{{(l)}^{-1}}{{b}_{i}}(l)<=1$}\\

Where, $N$ are the total number of users in the system, $r$ are the number of servers in the system and 
$\alpha$ is the tuning parameter.
The formulation is in the standard form and can be solved using MATLAB GP solver. For application specific requirements, we can also use additional constraints of minimum and maximum delays.

\subsection{Interpreting Problem}
The problem can be interpreted as follows:
The objective function is \[{{f}_{0}}(x)=\sum\limits_{l=1}^{N}{\alpha d(l)c(l)s(l)b{{(l)}^{-1}}}\] and the constraints are ${{f}_{i}}(x)\le {{u}_{i}}$ which are all posynomials. Since for ${{u}_{i}}=1$, the problem reduces to original GP and we can denote the perturbed problem as $p(\textbf{1})$ where \textbf{1} is the vector of all ones for all the constraints. If ${{u}_{i}}>1$, the constraint is loosened and therefore we may say that the $i$th inequality is less in effect. For example, we may loosen the delay constraints and the system would be able to accommodate more users in it. On the contrary ${{u}_{i}}<1$ would tighten the constraints, for instance, if we tighten the load constraint, the system would support less number of users. In any case, the variability of these constraints is important for a real-time network because a real network is sensitive to these changes. A detailed perturbation and sensitivity analysis is therefore studied in this work and is required for an iterative distributed framework which will be time-variant.
\section{Data Collection}
As we know, data measurement is a prime factor in accurate modeling, we use the network shown in Fig. 2 to collect the real-time data. All the nodes are virtual machines (VMs) located at Orbit Lab [4]. The router instances are click-based routers [5] while the client and servers are VMs running linux Ubuntu 14.04. Each of these machines run an instance of client or server and can communicate (\textit{ping} command) with each other.

\begin{figure}[thpb]
\centering
\includegraphics[width=0.3\textwidth]{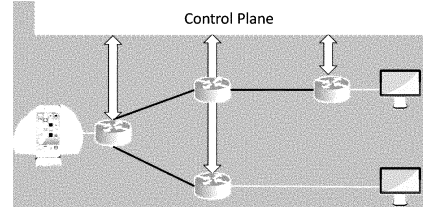}
\caption{Sample Network for Data Collection}
\label{fig:network}
\end{figure}

The server loads are emulated using varying delay in processing requests by the client while the energy is considered to be proportional to the number of hops. It is to be noted counter-intuitively that the client doesn't necessarily connects to the bottom server in spite of being reached by lower number of hops as server load plays an important metric. The results and discussion are presented in the next section.

\section{Results and Analysis}
We consider two routes as per Fig. 2, one using top three routers with varying server load and second using the bottom two routers. Fig. 3 shows the usage of convex optimization solved Application Specific Routing metric to select the optimal server reached by best path. Both the path provides a metric which is the objective function for us and we choose the minimal value of such a metric to decide a server and the path jointly. It is to be noted that a single server can be reached via multiple paths in real networks and therefore the joint optimization considers following (a) path to be chosen at each router, (b) server to be selected to reach the optimization and (c) constraints to be obeyed for optimization energy of the network and other network-level parameters such as delay, bandwidth etc.

\begin{figure}[thpb]
\centering
\includegraphics[width=0.5\textwidth]{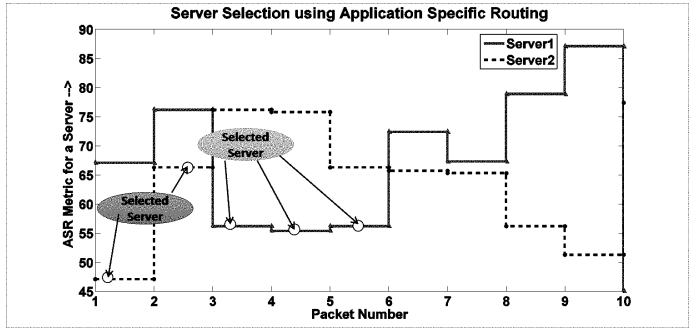}
\caption{Server Selection using Application Specific Routing Metric}
\label{fig:ss}
\end{figure}

Table \ref{tab:energy} highlights the energy consumed at each router after the optimization. The data is an important metric for the joint optimization and therefore is presented here. It can be noted that route 2 consists of only two routers (1 and 4) and therefore the sum of energies of these two routers is maximum.

\begin{table}[h]
\caption{Total Energy Consumed (\textit{in units})}
\label{tab:energy}
\begin{center}
\begin{tabular}{|c||c|}
\hline
Router Number & Energy\\
\hline
Router 1 & 783\\
\hline
Router 2 & 670\\
\hline
Router 3 & 674\\
\hline
Router 4 & 696\\
\hline
\end{tabular}
\end{center}
\end{table}

Figure 4 details the load distributed on both the servers using the application specific routing. It is interesting to note that there is near-equal distribution of load on the servers which is a positive metric for any system. The imbalance on loads due to the client running the application is due to the fact that there are different emulations on each of the servers at different point in time. It is also observed that the system is limited by different link bandwidths at different packets which is the minimum of all the bandwidths for a path.

\begin{figure}[thpb]
\centering
\includegraphics[width=0.5\textwidth]{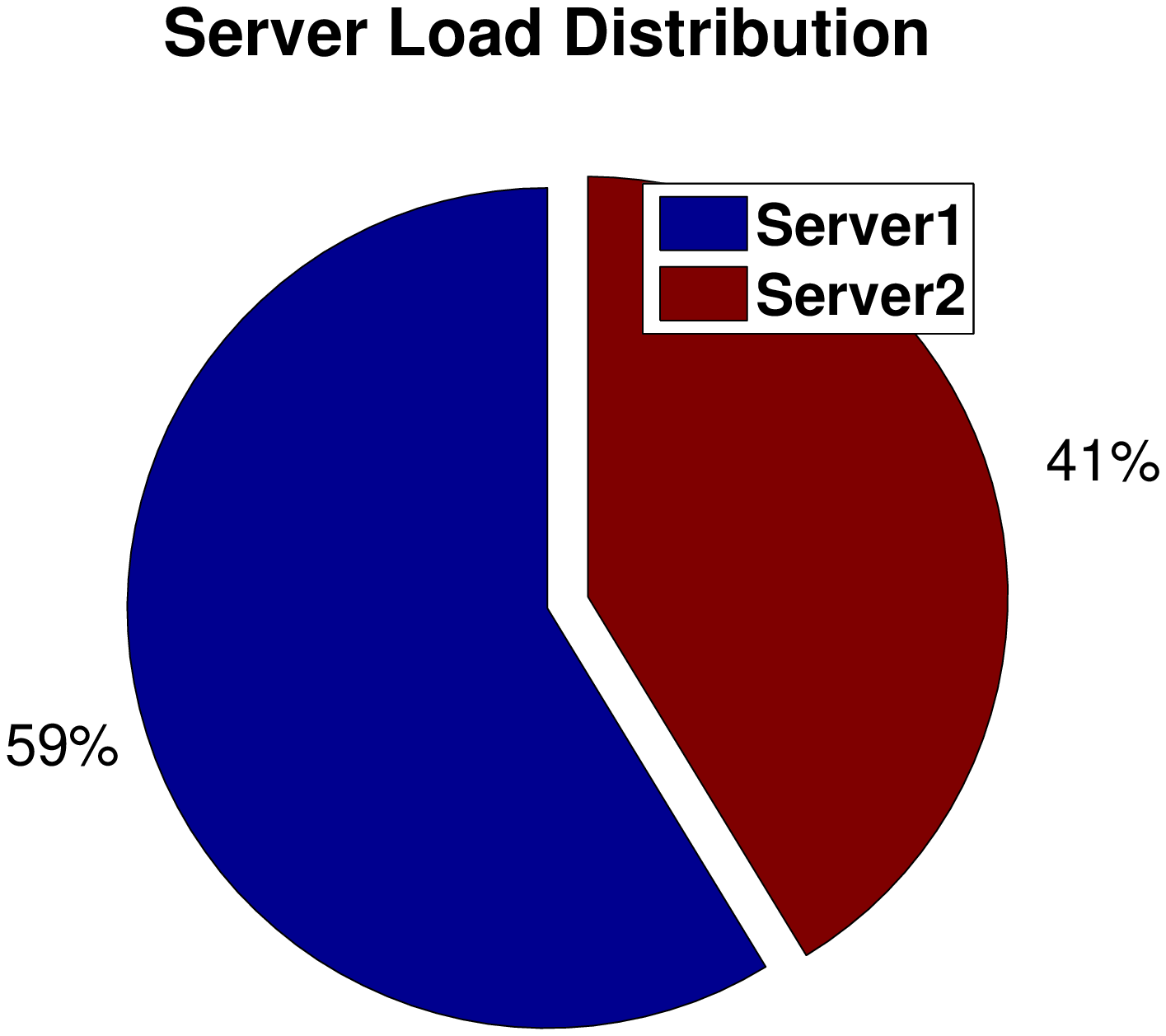}
\caption{Server Load Distribution}
\label{fig:load}
\end{figure}

Figure 5 plots the varied delay at each router in our network. It is worth noticing that the delay is highly variable and if is not for joint optimization, it is impossible to arrive at a consensus on path and server for a given packet. Also, it is worthwhile to mention here that the routers in our network maintains two metrics to store the delay: (i) long term estimate and (ii) current value. The former is a moving average which provides us the overall reliability of a path while the later provides the current status of the path. In general, if long term delay is more than short term delay, the packet can be marked for expedited forwarding while if short term delay is more than long term delay, the packet can be moved to a deferred queue and can be processed a little later when the path metric improves. 

\begin{figure}[thpb]
\centering
\includegraphics[width=0.5\textwidth]{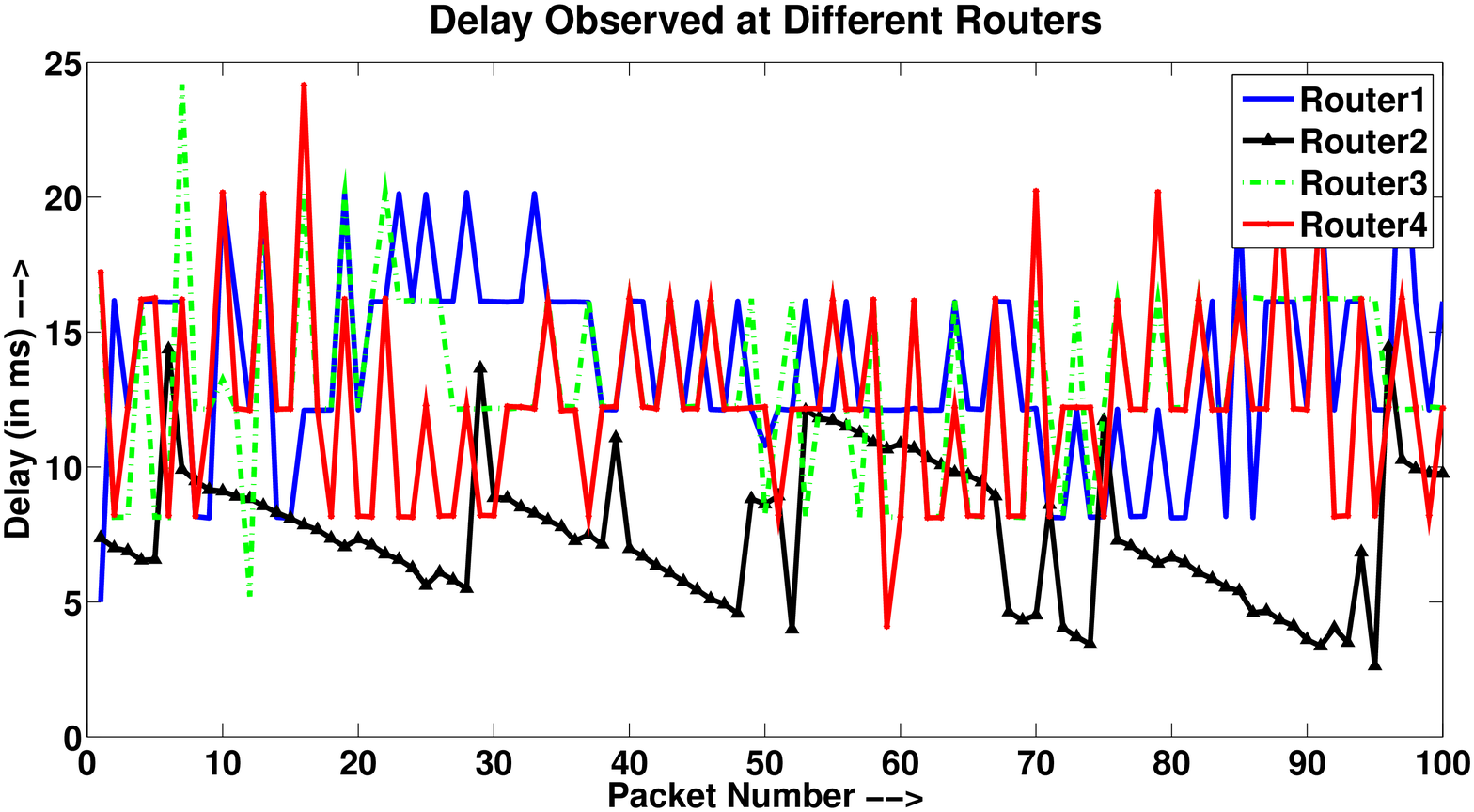}
\caption{Server Load Distribution}
\label{fig:delay}
\end{figure}

In our design, the value of alpha in the problem formulation is chosen for two particular reasons: (a) It allows us to provide a normalized input to the solver and (b) By varying the value of alpha, we can make sure that the solution converges and does not go to unbounded state.

As discussed earlier, the perturbation plays an important role in networking applications as the resources do run out or become excessive at each iterations, we did the perturbation analysis by tightening the constraints. We observed that the energy consumption varied widely as shown in Table \ref{tab:energy1}. The data suggests that on tightening the constraints, route 2 becomes less preferred if energy is considered to be prime goal. This can be ascertained by looking at the sum of energies of router 1 and router 4 which belong to route or path 2. 

\begin{table}[h]
\caption{Total Energy Consumed in the Perturbed Problem (\textit{in units})}
\label{tab:energy1}
\begin{center}
\begin{tabular}{|c||c|}
\hline
Router Number & Energy\\
\hline
Router 1 & 790\\
\hline
Router 2 & 659\\
\hline
Router 3 & 663\\
\hline
Router 4 & 750\\
\hline
\end{tabular}
\end{center}
\end{table}

As we can observe in Fig. 6, router 1 (also server 1 in this case), is selected 80\% of time while server 2 is selected only for 20\% of time when the constraints are tightened. This provides us with an interesting insight about the network that when the constraints are tightened in the network optimization, the loaded network becomes more loaded in order to satisfy the constraints. This is similar to TCP congestion control problem where more congestion leads to packet drop and therefore more retransmissions and more congestion. The study presented in this work focused mainly on the optimizing network path using a central controller and a combined metric of multiple parameters. We can also explore to use a few of the selected parameters at selected time durations for example during the day time we can only work on delay and throughput while during the night when there are less users we can use energy conservation as primary goal. This will lead to a multi-objective joint optimization for our future work.

\begin{figure}[thpb]
\centering
\includegraphics[width=0.5\textwidth]{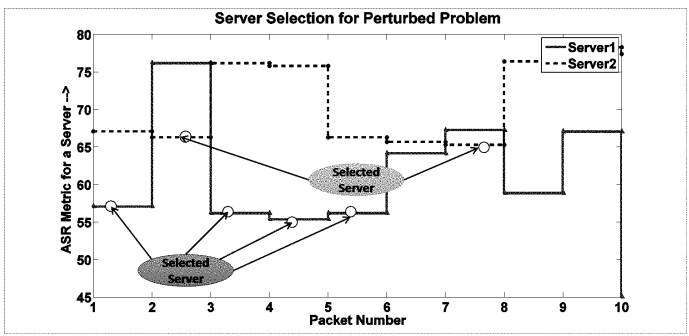}
\caption{Server Selection for Perturbed Problem}
\label{fig:per}
\end{figure}

\section{CONCLUSION}
This project accomplishes to choose an optimal path for application specific routing using a GP convex solver in an anycast network for multiple users. The method is illustrated with supporting results while using a central controller. In order to make this model realistic, we will used an iteration based distributed model in future and also a time-based multiple objective function method for better overall performance of a network.

\section*{ACKNOWLEDGMENT}
I would like to acknowledge Prof. Kristin Dana for her valuable inputs and support throughout the semester.

\end{document}